\documentclass[12pt]{iopart}
\usepackage{graphicx}

\begin{document}

\title[Vortex Tubes of Turbulent Solar Convection]{Vortex Tubes of Turbulent Solar Convection}

\author{I.N. Kitiashvili$^{1-4}$, A.G. Kosovichev$^1$, N.N. Mansour$^5$,\\S.~K. Lele$^{2,6}$ and A.A. Wray$^5$}
\address{$^1$ W.W. Hansen Experimental Physics Laboratory, Stanford University, Stanford, CA 94305, USA}
\address{$^2$Center for Turbulence Research, Stanford University, Stanford, CA 94305, USA}
\address{$^3$ NORDITA, Stockholm, SE-10691, Sweden}
\address{$^4$ Kazan Federal University, Kazan, 420008, Russia}
\address{$^5$ NASA Ames Research Center, Moffett Field, Mountain View, CA 94040, USA}
\address{$^6$Department of Mechanical Engineering, Stanford University, Stanford, CA 94305, USA}
\ead{irinasun@stanford.edu}

\begin{abstract}
Investigation of the turbulent properties of solar convection is extremely important for understanding the multi-scale dynamics observed on the solar surface. In particular, recent high-resolution observations have revealed ubiquitous vortical structures, and numerical simulations have demonstrated links between vortex tube dynamics and magnetic field organization and have shown the importance of vortex tube interactions in the mechanisms of acoustic wave excitation on the Sun.
In this paper we investigate the mechanisms of the formation of vortex tubes in highly-turbulent convective flows near the solar surface by using realistic radiative hydrodynamic LES simulations. Analysis of data from the simulations indicates two basic processes of vortex tube formation: 1) development of small-scale convective instability inside convective granules, and 2) a Kelvin-Helmholtz type instability of shearing flows in intergranular lanes. Our analysis shows that vortex stretching during these processes is a primary source of generation of small-scale vorticity on the Sun.
\end{abstract}
\maketitle

\section{Introduction}
Turbulent convection is a source of various phenomena observed on the Sun, such as coronal mass ejections, self-organized magnetic structures (that appear on the surface as sunspots and pores), and other non-linear multi-scale dynamical structures and phenomena. Modern realistic numerical simulations of solar turbulent phenomena are based on physical first principles and take into account the real-gas equation of state, radiative transfer, chemical composition, and the effects of magnetic fields. Realistic 3D numerical simulations have reproduced and explained many observed effects in sunspots and magnetic active regions \cite{kiti2009,kiti2010b,rempel2011,stein2011} and in quiet-Sun regions \cite{stein2000,steiner2010,kiti2011}. Thus, numerical simulations provide important insights into the physical mechanisms of solar phenomena.

  Vorticity is one of the basic properties of turbulent flows. Therefore it is not surprising that swirling motions are found in observations of the highly-turbulent solar magnetoconvection. Large-scale vortical behavior, once evidenced by sunspot rotation, was first observed on the Sun by Secchi (1857). Later, vortex flows in the photosphere ($\sim 3$ Mm in diameter) were detected by Brandt et al. \cite{brandt1988}, and small-scale swirling flows ($\sim 0.5$ Mm) were observed by Wang et al. \cite{wang1995}.

 Observations have shown that vortices are ubiquitous in non-magnetic quiet-Sun regions \cite{potzi2005,bonet08,bonet10} and are associated with intergranular lanes. Such swirling motions correspond to vertically oriented vortex tube structures that have been found in numerical simulations \cite{stein2000,kiti2011,brand1996}. The numerical simulations have demonstrated important links between vortex tube dynamics and processes of magnetic self-organization \cite{kiti2010b} and also with acoustic wave excitation \cite{kiti2011} and have shown that vortex tubes play a fundamental role in solar magnetic flux dynamics. Also, small-scale horizontal vortex tubes located along granule edges were found both in numerical simulations and in observations with the balloon observatory SUNRISE \cite{steiner2010}. Recently, vortical motions were observed inside granules by the NST/BBSO telescope (P. Goode, private communication). However, even the highest resolution observations are not capable of resolving the internal structure of the vortex tubes. Thus, it is important to investigate in detail the mechanism of vortex formation and dynamics using high-resolution numerical simulations.

Vortex tube formation can occur due to different processes. According to our numerical simulations, generation of vortex tubes on the Sun can be driven by the development of convective instabilities, accompanied by emergence of small-scale plumes inside granules or granule splitting, and also by Kelvin-Helmholtz instability of shearing flows in the intergranular lanes. Both vertical and horizontal vortex tubes are generated, but the vertical tubes predominate, as the horizontal tubes are often dragged into intergranular lanes where they become vertical due to surrounding downflows.

\section{Computational setup}

For this investigation of vortical structures in the turbulent convective near-surface boundary layer of the Sun we use the 3D radiative MHD `SolarBox' code developed at the NASA/Ames Research Center and the Stanford Center for Turbulence Research by Alan Wray and his colleagues \cite{jacoutot08a}. The code takes into account fluid flow compressibility in a highly stratified medium, the real-gas equation of state, the standard model of the solar interior \cite{chris1996}, and the OPAL opacity tables. Radiative transfer between fluid elements is calculated using a 3D multi-spectral-bin method assuming local thermodynamic equilibrium. The physical description of the dynamical properties of solar convection was improved through implementation of subgrid-scale turbulence models, which effectively increases the Reynolds number and provides representation of small-scale motions closer to the reality. This approach, based on Large-Eddy Simulation (LES) models of turbulence, has demonstrated good agreement between numerical modeling and observations \cite{jacoutot08a}.

For the current study, the simulation results have been obtained for two different computational domains: $6.4\times6.4\times5.5$~Mm$^3$ (with a 1~Mm layer of the atmosphere) and $3.2\times3.2\times7.5$~Mm$^3$ (with a 2~Mm atmospheric layer), and various grid resolutions: 12.5~km and 6.25~km in the horizontal direction, and 10~km and 6~km in the vertical direction. The lateral boundary conditions are periodic. The top boundary is open to mass, momentum and energy transfers, and also to radiative flux. The bottom boundary is open only for radiation and simulates energy input from the interior of the Sun. These high-resolution simulations reveal details of vortex tube formation and dynamics in solar granular convection, as described in the following sections.

\section{Vortex tubes formation by convective granular instability}

\begin{figure}
\begin{center}
\includegraphics[width=1\linewidth]{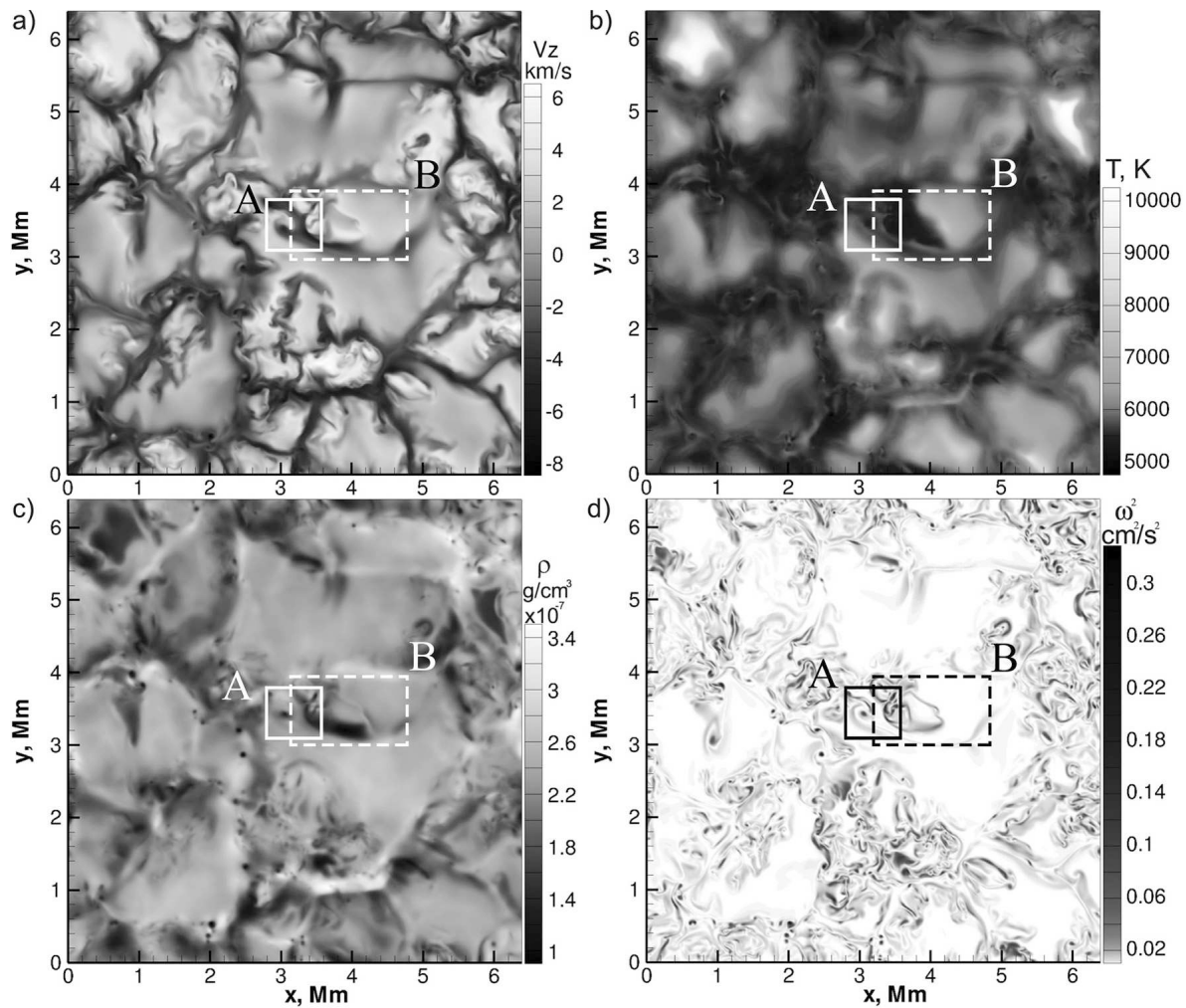}
\end{center}
\caption{Snapshots of a surface layer illustrate distributions of ({\it a}) vertical velocity, ({\it b}) temperature, ({\it c}) density, and ({\it d}) enstrophy. The rectangular regions indicate places of formation of a vortex tube from the granular instabilities due to 1) a local upflow in region `A' (solid rectangle; this region is shown in detail Fig.~\ref{fig:snapshots},~\ref{fig:enstrophy}); and 2) splitting of granule, region `B' (dashed rectangle; analyzed in detail in Fig.~\ref{fig:shallow},~\ref{fig:shallow3D}). \label{fig:snapshots-tot}}
\end{figure}

Subsurface layers of the Sun are highly turbulent and inhomogeneous. According to observations, the location of swirling motions (interpreted as vertical vortex tubes) is mostly associated with the intergranular lanes \cite{potzi2005,bonet08}. Similar association was found in numerical simulations \cite{stein2000,kiti2011,stein1998}. The numerical simulations presented in this paper show that vertical vortex tubes can also be formed inside granules because of a local instability caused by small-scale upflowing plumes. Also, splitting of granules due to convective instability can produce ``cookie cutter" structures at granular edges, which perturb flows at the granule top boundary and can produce vortices. In Fig.~\ref{fig:snapshots-tot} we show a snapshot of the solar surface properties obtained in the simulations: vertical velocity (panel {\it a}), temperature ({\it b}), density ({\it c}), and enstrophy (panel {\it d}). The vortex tube structures are best visible in the density distribution (Fig.~\ref{fig:snapshots-tot}{\it c}) as low-density (dark) points in the intergranular lanes. For a detailed analysis we selected two small subregions, marked `A' and `B', to illustrate the typical vortex formation process.

\subsection{Upflow plumes inside granular eddies}
\begin{figure}
\begin{center}
\includegraphics[width=1\linewidth]{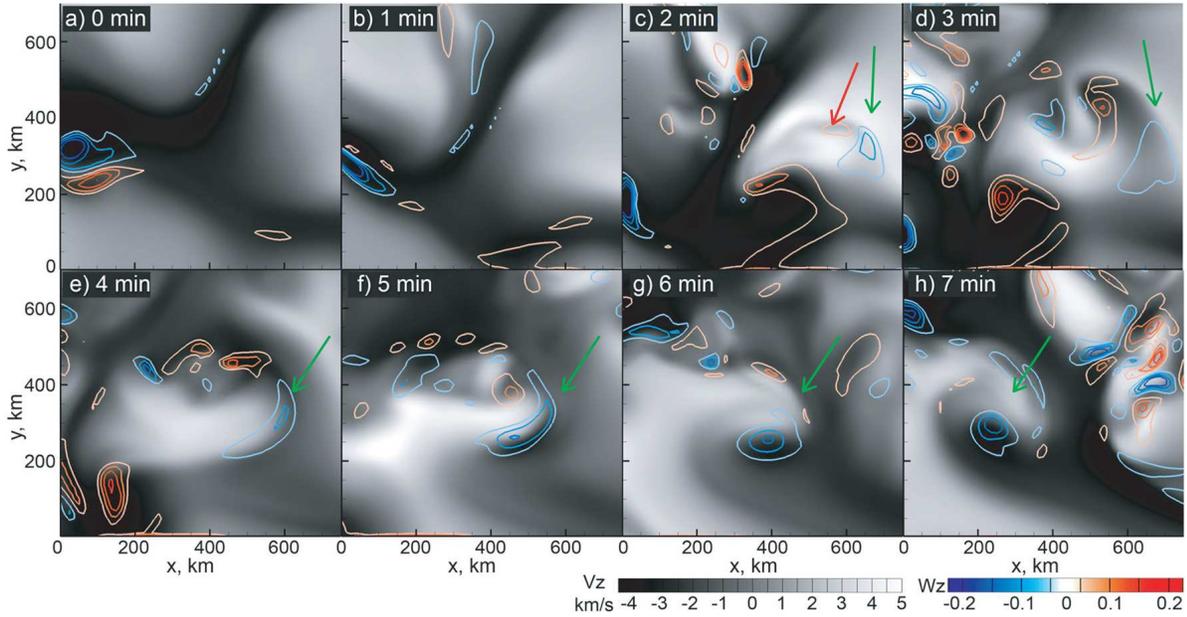}
\end{center}
\caption{Formation of a vortex tube on the solar surface in region `A' of Fig.~\ref{fig:snapshots-tot}. Grey-scale background shows the distribution of the vertical velocity, and contour lines show the vertical vorticity. This sequence illustrates the process of development of the vortical structure, which starts inside a granule. Red arrow indicates local upflow in the granule. Green arrows point out changes of the initial negative vorticity in part of vortex sheet (panels {\it c-d}) and in the vertical vortex tube (panels {\it g-h}). \label{fig:snapshots}}
\end{figure}

Vortex tube formation inside granules is a result of a complicated interaction of turbulent flows. In the surface layer, the development of convective instability initially represents a localized upflow in a granule (red arrow in Fig.~\ref{fig:snapshots}{\it a}), which is accompanied by a mixture of weak vortical motions on the surface (the vertical vorticity distribution in region `A' is shown by contour lines in Fig.~\ref{fig:snapshots}{\it a-c}). Overturning of this upflow plume increases the instability region and destroys the granule  (Fig.~\ref{fig:snapshots}{\it d-e}). Finally, the overturning vortical motions form a compact and relatively stable vortex tube (Fig.~\ref{fig:snapshots}{\it d-h}). The formation of the vortex tube and its development on the solar surface indicated by the green arrow. However, the vortical flows observed on the surface are only a part of complicated turbulent motions below the surface. In order to better understand the physics of the vortex development in convective granules we investigate the flow behavior in subsurface layers.

\begin{figure}
\begin{center}
\includegraphics[width=1\linewidth]{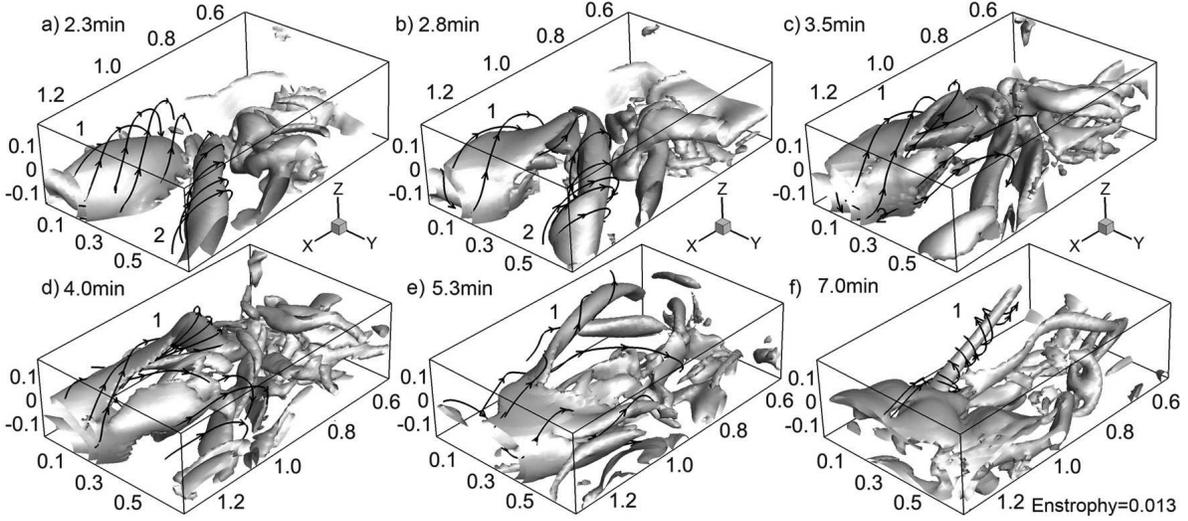}
\end{center}
\caption{Three-dimensional rendering of enstrophy in region `A' of Fig.~\ref{fig:snapshots-tot}, $(\nabla \times \vec{u})^{2} $, illustrating the evolution of helical structures. Deformation of the vortical structure $\sharp1$ into a sheet-like belt by strong upflows and its stretching by horizontal flows causes overturning of the structure's velocity gradients. Finally, this structure becomes more compact and evolves into a vertical vortex tube. The black streamlines illustrate the behavior of convective flows. Scales on the coordinate axes are in Mm. \label{fig:enstrophy}}
\end{figure}

The flow evolution of a typical granule revealed in our high-resolution simulations and illustrated in Figures 2 -- 3 is fairly complicated and can be described as follows. The diverging flows of a granule contain weak vortical motions along the granule edges (Fig.~\ref{fig:enstrophy}{\it a}). Enstrophy isosurfaces shown in Figure~\ref{fig:enstrophy}{\it a} indicate that at this stage the lengths of two vortex tubes (marked as structures $\sharp1$ and $\sharp2$) are $0.5 - 1$~Mm. During further evolution, surrounding flows compress the granule and transform the weakly helical structure $\sharp1$ into a vortex sheet stretched by horizontal diverging flows (Figs.~\ref{fig:enstrophy}{\it b-c}). Magnified by compression, the granular upflows carry the vortex into the upper layers. Also, surrounding shearing flows compress the horizontal vortex tube $\sharp2$  and can break it into small-scale vortical features (Figs.~\ref{fig:enstrophy}{\it c-f}).

In the subphotospheric layers the vortex sheet becomes unstable and splits into several segments (Figs.~\ref{fig:enstrophy}{\it c-d}). At the same time, the sheet-like part of structure $\sharp1$ starts overturning (Figs.~\ref{fig:enstrophy}{\it c-d}) because of the vertical velocity gradient between the middle region of the granule and its edge. In this case, the process of sheet overturning is magnified by the second vortical structure as shown on Fig.~\ref{fig:enstrophy}{\it b}, which is later destroyed by convection. During the overturning, the forming vortex tube is gradually moving into the intergranular lane, where surrounding downflows contribute to the overturning shearing motions and deform the sheet-like structure into an inclined vortex tube (Figs.~\ref{fig:enstrophy}{\it e-f}), which then becomes vertical.

\begin{figure}
\begin{center}
\includegraphics[width=1\linewidth]{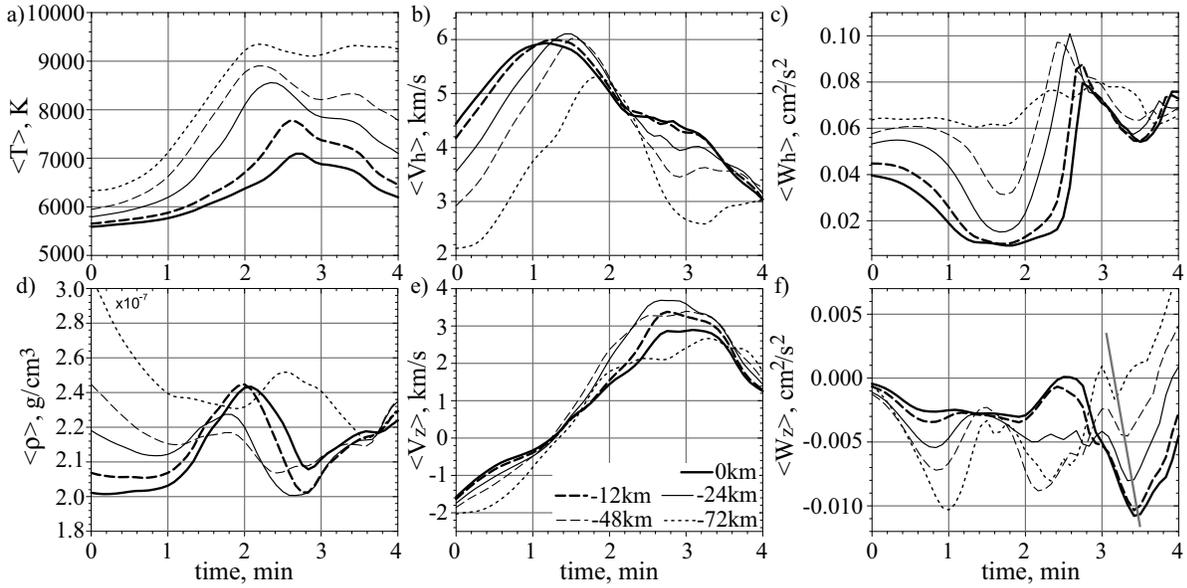}
\end{center}
\caption{Variations of mean temperature ({\it a}), horizontal velocity and vorticity ({\it b, c}), density ({\it d}), and vertical velocity and vorticity ({\it e, f}) for different depths in a region of vortex tube formation at the initial stage. The temperature and density anomalies (`bumps' on panels {\it a, d}), which migrate from deeper layers to the surface, are carried outward by the rising vortex tube. Velocity profiles illustrate the acceleration of horizontal flows due to compression of the granule by surrounding flows (panel {\it b}), and the transformation of the flows into upflow plumes ({\it e}). Comparison of the mean horizontal and vertical vorticities shows a correlation for the subphotospheric layers, where increasing vertical vorticity is accompanied by a decreasing horizontal component of vorticity (panels {\it c, f}). \label{fig:granule}}
\end{figure}

The overturning motions significantly affect the flow dynamics and thermodynamic properties in a local region of the instability and can be observed as fluctuations of mean quantities in the region. The typical flow speed in a granule is $\approx1$~km/s, but in the local upflows, the sizes of which are $\sim 100$~km, the flow speed can be $\sim 2$~km/s, and the temperature can be higher by $\sim$~400~K. These upflows are accompanied by weak helical motions in the subphotospheric layers (Fig.~\ref{fig:granule}). Thus, in this case vortex tube formation is caused by a granular instability, accompanied by high-speed small-scale upflows inside granules, which form a vortex sheet and drag it from below to the surface.

Figure~\ref{fig:granule} shows the temporal evolution of temperature, density, velocity, and vorticity with time at different depths. The initially expanding and rising vortex sheet causes a local compression of the flow and a temperature increase (Fig.~\ref{fig:granule}{\it a, d}). Behavior of the velocity field with time at different depths shows an increase of velocity during compression of the granule by surrounding flows, followed by granule decay occuring during the overturning phase (Fig.~\ref{fig:granule}{\it b, e}). The interplay of the horizontal and vertical vorticity components shown in Fig.~\ref{fig:granule}{\it c, f} illustrates a partial conversion of horizontal vorticity into vertical vorticity during overturning of the horizontally oriented vortex sheet. The temporal evolution of the vertical vorticity shows that the formation of a vortex tube starts in subsurface layers. This is indicated by deep minima in the vertical vorticity at $t \sim 3.2$~min in Fig.~\ref{fig:granule}{\it f}, which extend with time toward the surface. This trend is shown by a grey line in Fig.~\ref{fig:granule}{\it f}.

In addition to the convective instability resulting in a strong localized upflow motion inside granules, spitting of granules is also observed in our simulations. The results show that this process also can be accompanied by the formation of vortex tubes. In the following subsection we consider this type of granular instability.

\subsection{Granule splitting}
\begin{figure}
\begin{center}
\includegraphics[width=1\linewidth]{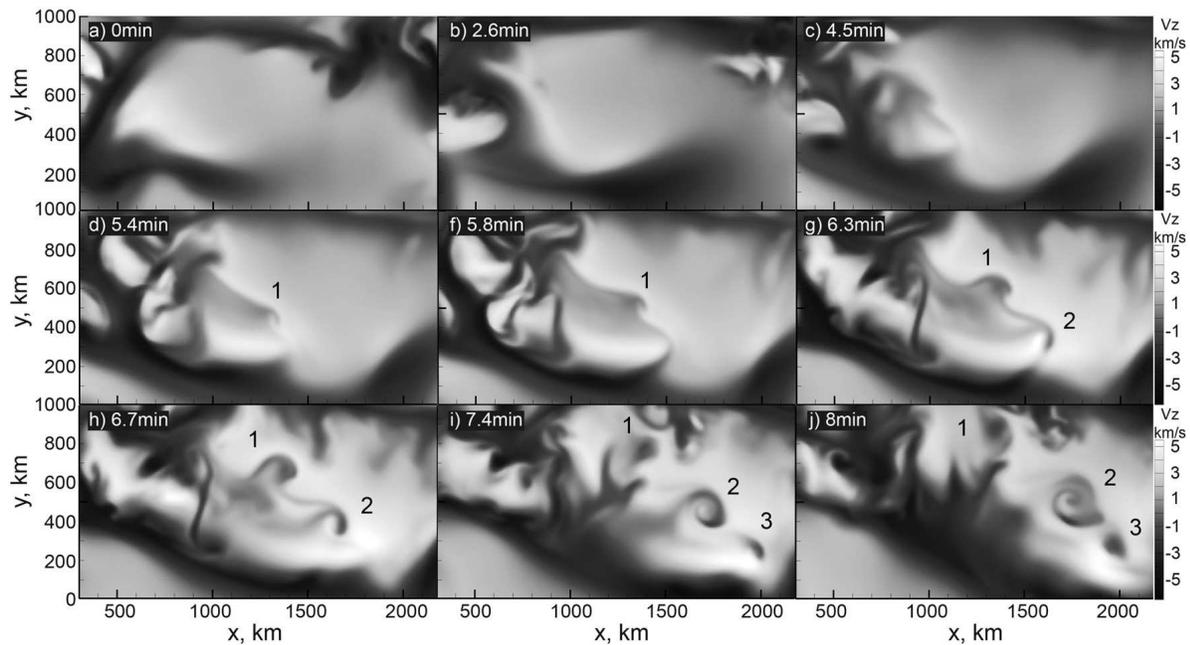}
\end{center}
\caption{Formation of a series of small-scale vortex tubes on the surface of a granule during splitting in region `B' of Fig.~\ref{fig:snapshots-tot}. Grey-scale images show the vertical velocity distribution on the surface. \label{fig:shallow}}
\end{figure}

Splitting of granules has been previously observed with different instruments \cite{kitai1979,roudier2003} and also in numerical simulations \cite{stein1998}. Our numerical simulations reveal a process of formation of relatively short-lived vortices inside granules during the splitting process. Figure~\ref{fig:shallow} shows a time sequence of the vertical velocity on the solar surface, where splitting of a granule is accompanied by a new shearing flow, which produces vortices. Actually, the initial stage of vortex formation is similar to the process described in Section 3.1. The main difference in the vortex formation process during granule splitting is that the vortices are continuously formed inside granules, without a tendency of moving into the intergranular lanes.

\begin{figure}
\begin{center}
\includegraphics[width=1\linewidth]{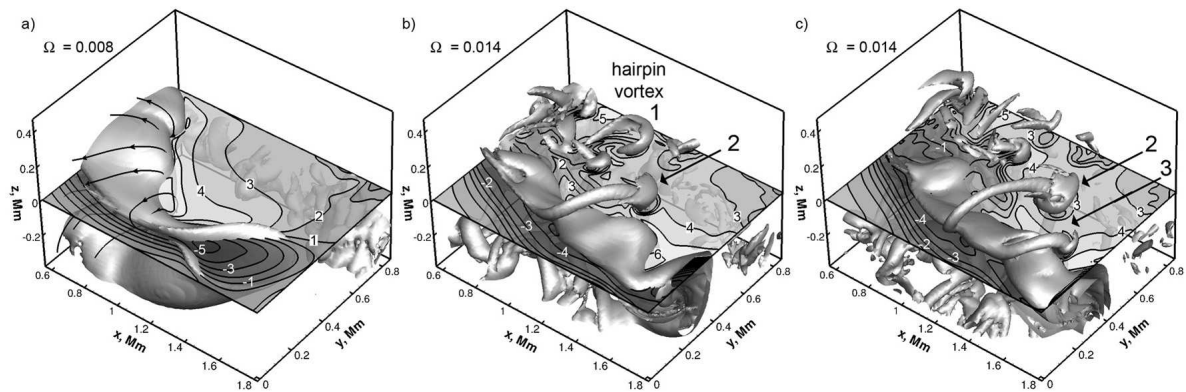}
\end{center}
\caption{Snapshots of the 3D structure of enstrophy (isosurfaces) illustrate different stages of vortex formation in region `B' of Fig.~\ref{fig:snapshots-tot}. The horizontal plane shows the vertical velocity distribution on the surface; isocontours indicate the magnitude of the velocity. \label{fig:shallow3D}}
\end{figure}

In this case, the shearing flows and later a deformed vortex sheet appear on the surface as a wavy pattern (or ``cake cutter" structure, Fig.~\ref{fig:shallow}). Each `wave' represents a horizontal cut through a vortex tube at the solar surface. In this case, three vortex tubes can be identified. In Figure~\ref{fig:shallow} we marked each vortex in the order of their formation. Figure~\ref{fig:shallow3D} illustrates the evolution of enstrophy (shown by isosurfaces) below and above the solar surface layer (horizontal plane). The process of rising and overturning of the vortex sheet (Fig.~\ref{fig:shallow3D}{\it a}) produces the `wavy' pattern in Figure~\ref{fig:shallow}{\it f}. During expansion and splitting, the vortex sheet initially forms vortex $\sharp1$, then vortex tube $\sharp2$ (Figs.~\ref{fig:shallow}{\it g} and ~\ref{fig:shallow3D}{\it b}), and, finally, vortex $\sharp3$ (Figs.~\ref{fig:shallow}{\it j} and~\ref{fig:shallow3D}{\it c}).

The first vortex has a loop-like (or `hairpin') structure. The lifetime of the `hairpin' vortex is very short, $\sim 5$~min and depends mostly on the magnitude of the diverging flows. The simulations show that such loop-like vortical structures are sensitive to surrounding flows, and during the evolution they are stretched, split by diverging flows, and diffuse. The vortex, $\sharp2$, is almost vertically oriented and has a very compact circular structure below and above the surface with diameter about 100~km. Because of diverging flows this vortex tube expands in diameter, up to $\sim 200$~km where the diverging flows are strongest (Fig.~\ref{fig:shallow3D}{\it b, c}). In our example, the expanded part of the vortex tube $\sharp2$ transforms into a ring-like substructure, which is separated from the initial vortex and finally destroyed by the diverging flows. This vortex tube $\sharp2$ is rather stable, with a lifetime up to 10~min, and almost all the time is connected to the `parent' vortex sheet. Vortex $\sharp3$ has a lifetime of few minutes and is formed during the decay of the `parent' vortex sheet (Figs.~\ref{fig:shallow}{\it j},~\ref{fig:shallow3D}{\it c}).

According to the numerical simulation, most of the vortex tubes are located in the intergranular lanes. Vortex tube formation caused by granular instabilities can be understood as a result of overturning of a vortex-sheet structure in upflowing plumes inside granules. This mechanism can be realized by small-scale and high-speed local upflows and during granule splitting. In those cases when the formation of vortices is initialized by local upflows inside granules, diverging granular flows move these vortices into intergranular lanes where they form stable and relatively long-lived vortex tubes. These series of vortex tubes, formed  during splitting of granule from a `cake cutter' structure, have short lifetimes, usually less than 10~min. Therefore the mechanism of the vortex formation by local upflows is primary, in the sense that it produces more stable and longer-living vortex tubes than granular splitting. However, our simulations show that such instabilities are not sufficiently frequent to explain the ubiquitous distribution of vortex tubes in the intergranular lanes. In the next section we consider vortex formation caused by shearing intergranular flows, which is probably the main mechanism of vortex tube formation.

\section{Kelvin-Helmholtz instability in intergranular lanes}

\begin{figure}
\begin{center}
\includegraphics[width=1\linewidth]{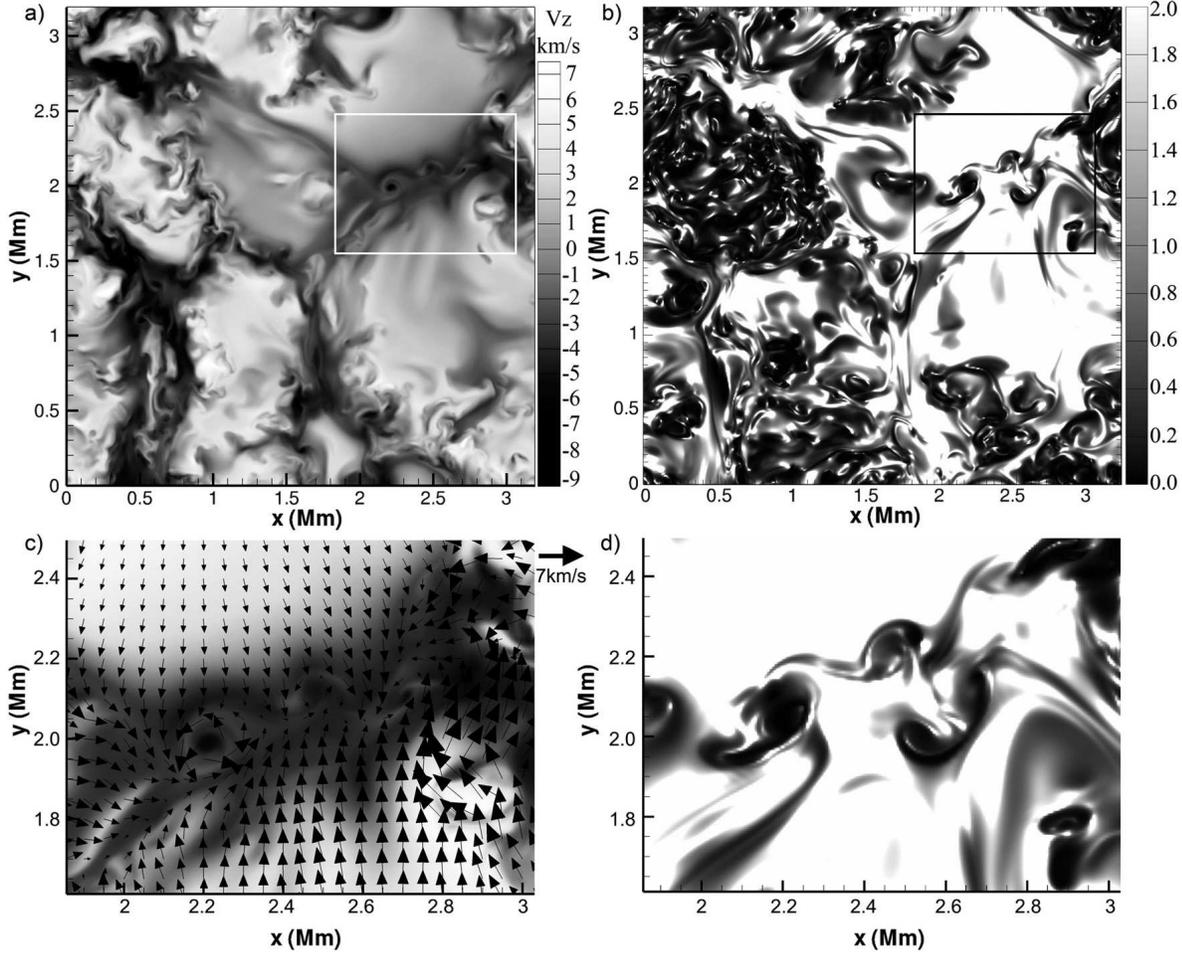}
\end{center}
\caption{Distribution of vertical velocity (panel {\it a}) and Richardson number (panel {\it b}) on the solar surface show an example of vortex tube formation in an intergranular lane due to the Kelvin-Helmholtz instability in high-resolution simulations (6.25~km step). Panels {\it c} and {\it d} show the distribution of vertical velocity and Richardson number in the rectangular subregions of panels {\it a} and {\it b} . Arrows show the horizontal velocity field. \label{fig:KH}}
\end{figure}

The formation of vortex tubes is most common in intergranular lanes, where strong downflows and horizontal shearing flows along the edges of granules are present. In this case, formation of one or several vortex tubes by shearing flows is due to the Kelvin-Helmholtz instability. Figure~\ref{fig:KH}{\it a} shows the vertical velocity distribution on the solar surface obtained from our high-resolution simulations with a 6.25~km grid interval. An example of an area where the Kelvin-Helmholtz instability develops is indicated by the rectangle. A similar flow instability can often be observed in the intergranular lanes on smaller scales.

The regions of development of the Kelvin-Helmholtz instability can be determined from the distribution of the Richardson number, $Ri=N^2/\left(\frac{d u_h}{dz}\right)^{2}$, where $N$ is the Brunt-V\"ais\"al\"a frequency and $u_h$ is the horizontal velocity. Usually estimates of the Richardson number can indicate transition from laminar to turbulent flow and vice versa. In the solar convection case characterized by very high Reynolds number ($\sim 10^{12}$), the Richardson number indicates a relative level of turbulence. Thus, the distribution of the Richardson number on the solar surface (Fig.~\ref{fig:KH}{\it b}) qualitatively indicates, where it is small, the predominant regions of formation of vortices.

In particular, the apparent decrease of Richardson number in the intergranular lanes corresponds to transition to more a turbulent regime, where horizontal shearing flows of different scales can initialize vortices (Fig.~\ref{fig:KH}{\it c}). Such shearing flows often have a complicated structure containing various flow streams propagating in different directions, strong density stratification, and gradients in the downflow speed.

\section{Enstrophy balance and sources of vorticity}

\begin{figure}
\begin{center}
\includegraphics[width=1\linewidth]{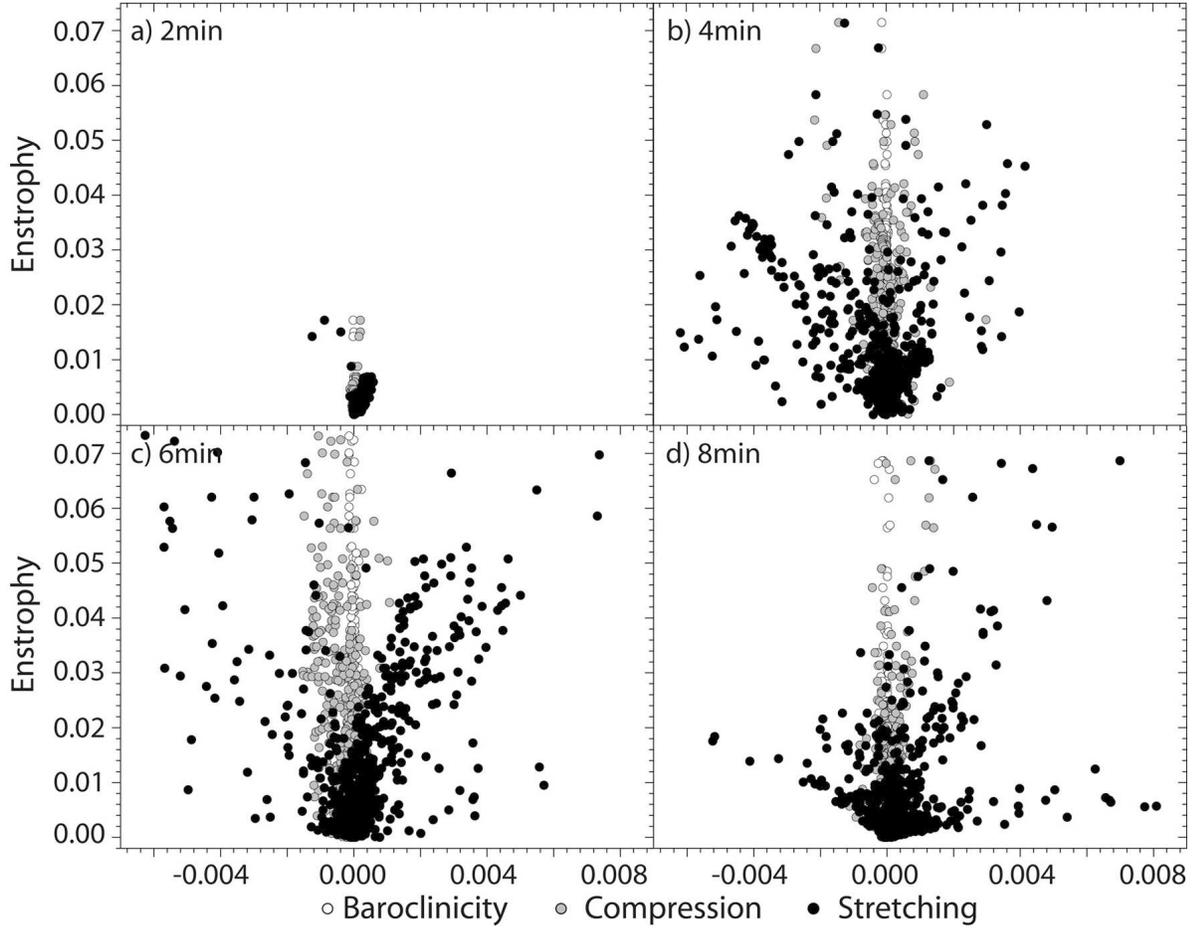}
\end{center}
\caption{Dependence of enstrophy on the three types of vorticity sources: baroclinicity (empty circles), compression (grey), and stretching (black circles), in a near-surface layer for different moments of time. \label{fig:En-BSC}}
\end{figure}

To investigate the process of the vortex formation in terms of the enstrophy evolution, we compare contributions of various vorticity sources, such as baloclinicity ($B$), compression ($C$), and stretching ($S$). The evolution of the enstrophy, $\omega^2$, is given by the following equation \cite{porter2000}
\begin{eqnarray}
    \frac{d\omega^2}{dt}=B+S+C,
\end{eqnarray}
where
\begin{eqnarray}
    B=2 \vec{\omega} \cdot \nabla P\times\nabla\frac{1}{\rho}, \qquad S=2\vec{\omega}\cdot \vec{\omega}\cdot\nabla \vec{u},  \qquad C=-2\omega^2\nabla\cdot\vec{u},
\end{eqnarray}
here $\vec\omega$ is vorticity, $P$ is gas pressure, $\rho$ is density, and $\vec u$ is flow velocity. The terms $B$, $S$, and $C$ represent the vorticity sources corresponding to baroclinicity, vorticity stretching, and compression respectively.

\begin{figure}
\begin{center}
\includegraphics[width=1\linewidth]{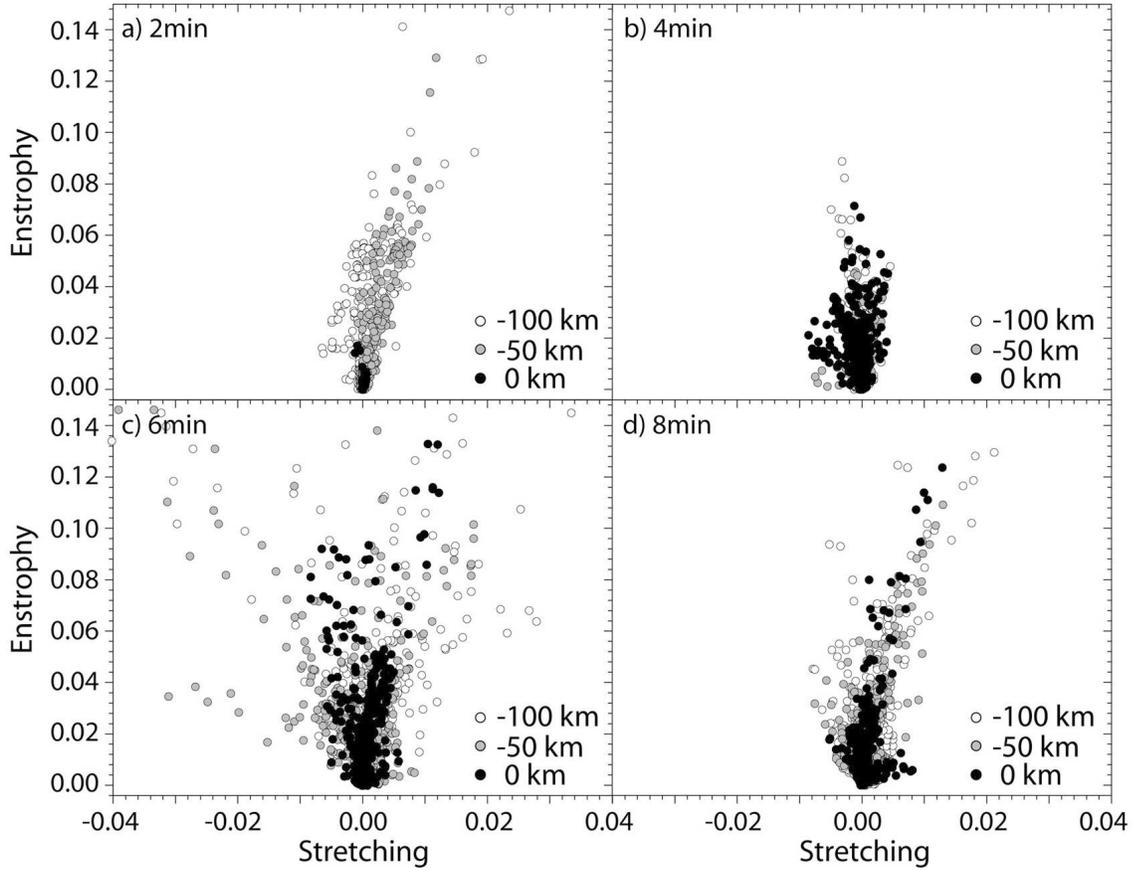}
\end{center}
\caption{Evolution of enstrophy vs. stretching for different stages of vortex formation at different depths: $-100$km (empty circles), $-50$km (grey), and surface layer (black circles). \label{fig:En-BSC_Depth}}
\end{figure}

Comparison of their contributions in the surface layer for different moments of time (Fig.~\ref{fig:En-BSC}) shows the dominant role of the vortex stretching effect. It is surprising that the baroclinicity effect does not play any significant role at this stage, but its contribution starts increasing when the vortex tube is almost formed. The variation of the stretching source with depth is significant only at a very initial stage of the vortex formation (Fig.~\ref{fig:En-BSC_Depth}{\it a}). During the vortex overturning stage the contribution of stretching to enstrophy production is similar at different depths (Fig.~\ref{fig:En-BSC_Depth}{\it b-c}) because of strong turbulent mixing. The subsequent expansion of the mixing region increases the influence of stretching, but a very complicated flow topology in this region makes the enstrophy dependence on the individual sources unclear. During the final stage of vortex formation the stretching is very consistent (in both magnitude and sign) at all layers, from the surface to 100~km below the surface, because the formed vertical vortex tube has almost the same flow topology at all these depths (Fig.~\ref{fig:En-BSC_Depth}{\it d}).

\begin{figure}
\begin{center}
\includegraphics[width=0.9\linewidth]{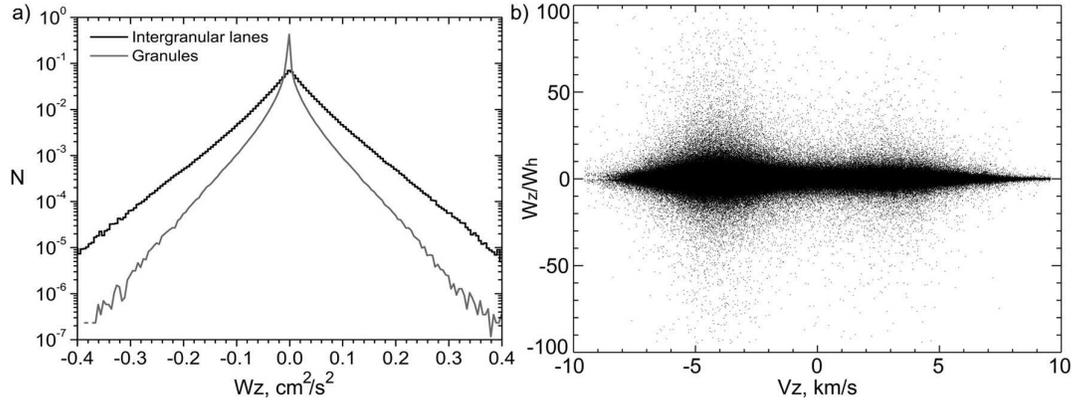}
\end{center}
\caption{Histograms of the distribution of vertical vorticity in intergranular lanes (black lane) and granules (grey lane), normalized by the total number of pixels for each region (panel {\it a}). The ratio of vertical and horizontal vorticity magnitudes vs. the vertical velocity component (panel {\it b}). \label{fig:hist}}
\end{figure}

The statistical distribution of the forming vortices is shown in Fig.~\ref{fig:hist}{\it a}, where black and grey curves correspond to the relative density of vorticity in intergranular lanes and in granules. The histograms are normalized by the total number of pixels of the developing vortices and were calculated for a near-surface $12.8\times12.8$~Mm$^2$ slice, with a 12.5~km/step resolution (or $1024^2$ grid-points), using a 15~min long dataset. Vertical vorticity of a relatively weak magnitude is dominant in granules and is mostly related to unstable small-scale fluctuations as discussed in Section 3. This also explains the dramatic decrease of the vorticity density in granules with increasing magnitude of vorticity (Fig.~\ref{fig:hist}{\it a}). In the intergranular lanes, the decrease of the vorticity density with the magnitude is less steep for two reasons: 1) the stable vortex tubes formed inside granules migrate into the intergranular lanes, and 2) strong vortices are formed there by shearing flows. The dependence of the ratio between the vertical, $\Omega_z$, and horizontal vorticity, $\Omega_h$, magnitudes on the local vertical velocity, Vz, shows concentration of vertical vortex tubes (where $\Omega_z/\Omega_h \gg 1$) in the intergranular lanes with vertical downward velocities of about 4~km/s, and a weaker increase of the number of vertical vortices in granules, associated with strong upflows (Fig.~\ref{fig:hist}{\it b}).

Vortex tube formation caused by vortex sheet overturning inside granules is a common phenomenon in our simulations. However, the distribution of vortices shows a clear tendency to accumulation in intergranular lanes. This can be explained by two effects: 1) the process of formation of vortex tubes is often accompanied by transport of new vortices into the intergranular lanes by diverging flows formed inside granules; and 2) preferential formation of vortices in the intergranular regions due to the Kelvin-Helmholtz instability in strong shearing flows.

\section{Discussion and conclusion}
Observations of turbulent solar convection have shown the existence of vortical motions on very different scales: from small-scale ($\sim 500$~km) to large-scale ($\sim 20,000$~km). The small-scale vortices, detected on the solar surface with large ground-based telescopes and balloon observations (SUNRISE, NST/BBSO), are an important part of the convective dynamics of the granulation layer. The topology of the turbulent flows can be very complicated and is often accompanied by arc-like vortical structures above the surface (Fig.~\ref{fig:shallow3D}{\it c}), which link the convective subsurface layers with the atmosphere. Therefore, a division of vortical structures into different types, e.g. horizontal and vertical vortex tubes, vortex sheets, etc., is only a simplification of the real picture of the turbulent flow.

 For this initial investigation we concentrated on purely hydrodynamic turbulent effects, although magnetic fields undoubtedly play a very important role in vortex tube formation. The simulation results have shown that vortex tubes are mostly concentrated in the intergranular lanes, but they are also formed inside the convective granules (Fig.~\ref{fig:snapshots-tot}). Vortex tube formation can be initiated by two basic processes: 1) small-scale convective instability leading to localized upflows inside granules, and 2) the Kelvin-Helmholtz instability of shearing flows in the intergranular lanes.

The convective granular instability usually results in small-scale local upflows inside granules, which initially form vortex sheets. Flow overturning in these sheets and their simultaneous advection into intergranular lanes produce vertically oriented vortex tubes (Figs.~\ref{fig:enstrophy},~\ref{fig:shallow3D}). It is interesting that a similar process occurring during splitting of granules can be a source of a series of different types of vortices (Fig.~\ref{fig:shallow}). The mechanism of vortex tube formation due to the Kelvin-Helmholtz instability (Fig.~\ref{fig:KH}{\it a}) works mostly in the intergranular lanes, where horizontal shearing flows are the strongest. The 3D topological structure of the shearing flows determines the dynamical properties of the vortices and their extension into the interior. The simulations reveal that vortex formation is a very common process on small scales not yet resolved in solar observations.

\ack{This work was partially supported by the NASA grant NNX10AC55G, the International Space Science Institute (Bern) and Nordita (Stockholm). The authors thank Douglas Gough, Phil Goode, Vasyl Yurchshin, Valentyna Abramenko, and participants of the Nordita and ISSI teams for interesting discussions and useful suggestions.}

\end{document}